\documentclass[12pt,aps,pra,preprint]{revtex4}
\usepackage{graphicx}
\usepackage{dcolumn}
\usepackage{amsmath}
\usepackage{amssymb}
\usepackage{float}
\usepackage{epstopdf}

\def\beq{\begin{equation}}
\def\eeq{\end{equation}}

\begin{document}
\title{Universality of the energy spectrum for two interacting harmonically trapped ultra-cold atoms in one and two dimensions}
\author{Aaron Farrell and Brandon P. van Zyl}
\affiliation{Department of Physics, St. Francis Xavier University, 
Antigonish, NS, Canada B2G 2W5} 
\date{\today}

\begin{abstract}
Motivated by the recent article of P.~Shea {\it et al.~} [Am.~J.~Phys.~ {\bf 77} (6), 2009]
we examine the exactly solvable problem of two harmonically trapped ultra-cold bosonic atoms interacting {\it via} a short range potential in one and two dimensions.
A straightforward application in one dimension shows that the
energy spectrum is universal, provided that the range of the potential is much smaller than the oscillator length, 
in addition to clearly illustrating why regularization is not required
in the limit of zero range.  The two dimensional problem is less trivial, requiring a more careful
treatment as compared to the one dimensional case.  Our two dimensional
analysis likewise reveals that the low-energy physics is also universal, in addition to providing a simple 
method for obtaining the appropriately regularized two dimensional pseudopotential.
\end{abstract}

\maketitle
\section{Introduction}
Recently, P.~Shea {\it et al.~}~\cite{shea2008} have discussed the problem of two bosonic atoms interacting {\it via} a
short range potential and trapped in a three dimensional (3D) spherically symmetric harmonic oscillator potential.  
Their work showed that the low energy properties of the 3D system are universal, irrespective of the
shape of the potential, provided the range is much smaller than the oscillator length.  In addition,
they developed the appropriate pseudopotential for a zero-range interaction in a manner
accessible to undergraduate students of physics with only an elementary knowledge of quantum mechanics
and scattering theory.  Specifically, no prior knowledge of self-adjoint extensions,
renormalization techniques, or dimensional regularization schemes are required to obtain equivalent results.

In this article, we present  details of the application of the techniques presented in Ref.~[\onlinecite{shea2008}] to both
one and two dimensional systems.
While the one dimensional (1D) problem proves
to be quite accessible, the two dimensional (2D) analogue turns out to be rather subtle. 
The current interest in the physics of
low-dimensional cold atom systems~\cite{lowdimensional} provides additional impetus for the results presented in this
paper.  In particular, it is now conceivable that analogous experiments to those performed in Ref.~[\onlinecite{stoferle}] can also be carried out on
low dimensional systems, in which case the universal aspects of the spectra derived here may be experimentally verified.

The plan for our paper is as follows.
In section II, we derive the pseudopotentials appropriate for a two-body interaction in the limit of zero range for both
the 1D and 2D systems.  Our approach clearly illustrates the concept of the pseudopotential
in the form of a regularized Dirac delta function while avoiding technical discussions about the self-adjointness of the two-body Hamiltonan, or of regularization operators
required to ensure the Hamiltonian's self-adjoint property.  Our result for the 1D pseudopotential scenario agrees with the literature,
whereas our 2D pseudopotential is ostensibly different from earlier published results.  Nevertheless, we argue that
our 2D pseudopotential is operationally equivalent provided that it acts upon the appropriate two
body wave function.  In section III, we show that the energy spectra in both the 1D and 2D systems are
universal and independent of the details of the pseudopotential provided that the range of the
interaction is much smaller than the oscillator length.  The universal properties for the energy spectrum we find in both 1D and 2D are
not well-known in the literature.~\cite{author_note2}   Our analysis also provides a sharp 
contrast to the results presented in 
Ref.~[\onlinecite{busch}] where the explicit properties of the regularized Dirac delta function are needed in order to obtain the spectrum.
In section IV, we finish with some concluding remarks and suggestions for future research in this area.

\section{The zero-range pseudopotential}
\subsection{One dimensional treatment}

We first consider a free system of two identical bosons, each of mass $M$, interacting
via a short-range symmetric potential in one dimension.  In the relative coordinate, 
$r=|r_1-r_2|\geq 0$, the $s$-state asymptotic scattering wave function is given by  
\begin{equation}\label{scattering1d}
\psi(x) \sim  \cos(k r + \delta(k))\ \ \ \ \ \ \ \ \ \ \ \  (r>b),
\end{equation}
where $b$ is the range of the interaction potential and $r \equiv |x|$.  Barlette
{\it et al.}~\cite{barlette2000} have already provided us with the
$s$-wave effective range expansion, relating the phase shift, $\delta(k)$, to the effective range, 
$r_0$, and scattering length, $a$, {\it viz.,} 
\begin{equation}\label{Erange1d}
k\tan(\delta(k)) = \frac{1}{a}+\frac{1}{2}r_0k^2+O(k^4)~,
\end{equation}
where the higher order terms  are shape dependent.~\cite{shea2008}
The effective range, $r_0$, is related to the range, $b$, in such a way that as $b\to0$, $r_0\to 0$. 
Thus, in the limit of zero-range, Eq.~(\ref{Erange1d}) reduces to $k\tan(\delta(k))= \frac{1}{a}$.  
Following Ref.~[\onlinecite{shea2008}], we extrapolate these results to bound states for positive $a$, where the $S$-matrix has poles at $\cot\delta = i$.
Utilizing the zero-range limit of Eq.~(\ref{Erange1d}), we immediately obtain $k= \frac{i}{a}$, which gives the bound state wave function,
\begin{equation}
\psi(x) = e^{-\frac{r}{a}}      \ \ \ \ \ \ \ \ \ \ \ \ \ \ \ \  (r>0),
\end{equation}
with binding energy $E= \hbar^2k^2/M=-\frac{\hbar^2}{Ma^2}$.  Equation (3) is exact for a zero-range potential and holds for $b \neq 0$
provided the size of the bound two-body system is much larger than the range of the potential responsible for the binding.  Under these
conditions, it is possible to construct an effective potential that reproduces the shape-independent results we have just obtained.
To this end, we consider the 1D Laplacian of Eq.~(3) with respect to the argument, and recall that $r\equiv |x|$.  Performing this operation
gives
  \begin{equation}\label{del1dWF}
\frac{d^2 \psi(x)}{dx^2} = \frac{1}{a^2}e^{-\frac{|x|}{a}}\left(\frac{d |x|}{dx}\right)^2-\frac{1}{a}e^{-\frac{|x|}{a}}\frac{d^2 |x|}{dx^2}.
\end{equation}
Using the relations $\left(\frac{d |x|}{dx}\right)^2=1$ and $\frac{d^2 |x|}{dx^2}=2\delta^{1d}(x)$ in Eq.~(\ref{del1dWF}) yields 
 \begin{equation}
-\frac{\hbar^2}{M}\frac{d^2 \psi(x)}{dx^2} -\frac{2\hbar^2}{Ma}\delta^{1d}(x)e^{-\frac{|x|}{a}} = -\frac{\hbar^2}{Ma^2}e^{-\frac{|x|}{a}},
\end{equation}
which, upon the substitutions $E=-\frac{\hbar^2}{Ma^2}$ and $\psi(x) = e^{-\frac{r}{a}}$,  reduces to
\begin{equation}\label{TISEVPP}
-\frac{\hbar^2}{M}\frac{d^2 \psi(x)}{dx^2} -\frac{2\hbar^2}{Ma}\delta^{1d}(x)\psi(x) = E\psi(x).
\end{equation}
Equation (\ref{TISEVPP}) it nothing more than the time-independent Schr\"odinger equation from which we see that the pseudopotential
\begin{equation}
V_{pp}^{1D}(x)= -\frac{2\hbar^2}{Ma}\delta^{1d}(x),
\end{equation}
reproduces the earlier results for the wave function and binding energy for any zero-range potential with a given scattering length.
This zero-range pseudopotential is in complete agreement with the literature,~\cite{busch, wodkiewicz, li} and naturally
 illustrates why no regularization of the Dirac delta function is required in one dimension.

\subsection{Two dimensional treatment}

We now move on to consider the same system as above, but now in strictly two dimensions.
In 2D the relative coordinate, $r=|r_1-r_2|$, $s$-state asymptotic scattering wave 
function is given by~\cite{Adhikari}  
\begin{equation}\label{E:WF1}
\psi(r)=\frac{u(r)}{\sqrt{r}} = \frac{\pi}{2}\left(\cot(\delta)J_0(k r)-N_0(k r)\right)\ \ \ \ \ \ \ \ \ \ \ \ (r>b)
\end{equation}
where $b$ is the range of the potential and $J_0(k r)$ and $N_0(k r)$ are, respectively, the zero order Bessel and Neumann functions.   Before we present our derivation of the 2D pseudopotential,
it is worthwhile clarifying our convention for the scattering length in two dimensions.  

In 1D and 3D systems, the definition of the $s$-wave scattering length is unambiguous.  Specifically, in the asymptotic region, and $E\to0$, the 1D and 3D problems reduce
to 
\begin{equation}
\frac{d^2}{dr^2}u(r)=0,
\end{equation}
where $u(r)=r\psi(r)$ in three dimensions and, letting $r=|x|$, $u(r)=\psi(r)$ in one dimension. This above 
equation is solved by $u(r)= C(r-a)$ where  $a$ is called the scattering length. 
In other words,  $a$ is identified as the intercept of the zero-energy wave function (or its
extrapolation) on the horizontal $r$-axis.  In 2D, we define $u(r)=\sqrt{r}\psi(r)$ and obtain the asymptotic,  reduced radial equation, 
\begin{equation}
\frac{d^2}{dr^2}u(r)+\frac{u(r)}{4r^2}=-k^2u(r).
\end{equation}
Now, as before, we let $k\to0$ and are required to solve the differential equation
\begin{equation}
\frac{d^2}{dr^2}u(r)+\frac{u(r)}{4r^2}=0~.
\end{equation}
The solution to Eq.~(11) is given by the function $u(r)=d_1\sqrt{r}\ln{r}+d_2\sqrt{r}$ 
where $d_1$ and $d_2$ are constants of integration.   Next, we rewrite our solution in the form
$u(r)=C\sqrt{r}\left(\ln{r}-\ln{a}\right)$.   Following the same arguments as for the 1D and 3D
cases, we define the scattering length as the  node in the asymptotic zero-energy wave function.  
Note that by definition, the scattering length in 2D is strictly positive, which is in stark contrast
to the 1D and 3D systems where $a$ can be of either sign.~\cite{shea2008}

Coming back to the problem at hand, we consider an interaction potential of zero-range and utilize the effective range expansion~\cite{verhaar}
\begin{equation}
\cot(\delta(k)) = \frac{2}{\pi}\left(\ln\left(\frac{ka}{2}\right) + \gamma \right)~,
\end{equation}
where $\gamma$ is the Euler constant, $\delta(k)$ is the scattering phase shift and $a$ is the previously defined scattering length. As before, we can extrapolate our results to a bound state.  
Thus we have $i= \frac{2}{\pi}\left(\ln\left(\frac{ka}{2}\right) + \gamma \right)$ which reduces to $k= \frac{2ie^{-\gamma}}{a}$.
Substitution of the latter expression into the scattering wave function, (\ref{E:WF1}), gives 
\begin{equation}
\psi(r) =  \frac{\pi}{2}\left(iJ_0\left(\frac{2ie^{-\gamma}r}{a}\right)-N_0\left(\frac{2ie^{-\gamma}r}{a}\right)\right) \ \ \ \ \ \ \ \ \ \ \ \ \ \ (r>0),
\end{equation}
The above equation can be re-written in terms of the modified Bessel functions, 
$I_0$ and $K_0$ by using the relations~\cite{handbook}
\begin{equation}\label{E:mod1}
J_0\left(iy\right)= I_0\left(y\right)~,
\end{equation}
and 
\begin{equation}\label{E:mod2}
N_0\left(iy\right)= iI_0\left(y\right)-\frac{2}{\pi}K_0\left(y\right)
\end{equation}
where $y$ is a generic argument. Using Eqs.~ (\ref{E:mod1}) and (\ref{E:mod2}) enables us to obtain the bound state
wave function, {\it viz.,}
\begin{equation}
\psi(r) = K_0\left(\frac{2e^{-\gamma}r}{a}\right) \ \ \ \ \ \ \ \ \ \ \ \ \ \ \ \ \\ \ \ \ \ \ \ \ \ \ \ \ \ \   (r>0),
\end{equation}
with binding energy $E = -\frac{4\hbar^{2}}{Ma^{2}e^{2\gamma}}$.   The appropriate 2D pseudopotential
can now be obtained by following the same arguments as for the 1D case.  Namely,
we first make the substitution $r=r(r)$, and then take the 2D Laplacian of our bound state wave function.  Some simple algebra gives
\begin{equation}\label{E:del1}
\nabla^{2}\psi(r)= \frac{2e^{-\gamma}}{a}\left(\frac{r'(r)}{r}+r''(r)\right)\left(K_0'\left(\frac{2e^{-\gamma}r(r)}{a}\right)\right)+\frac{4r'(r)^{2}e^{-2\gamma}}{a^{2}}\left(K_0''\left(\frac{2e^{-\gamma}r(r)}{a}\right)\right),
\end{equation}
where the primes denote derivatives with respect to the specific arguments.  We now make use
of the following useful properties of $K_0$,~\cite{handbook} 
\begin{equation}
K_0''\left(2 e^{-\gamma}\frac{r(r)}{a}\right)=K_0\left(2e^{-\gamma}\frac{r(r)}{a}\right)-\frac{a}{2e^{-\gamma}r(r)}
K_0'\left(2e^{-\gamma}\frac{r(r)}{a}\right),
\end{equation}
and
\begin{equation}
K_0'\left(2e^{-\gamma} \frac{r(r)}{a}\right)= \frac{a}{2e^{-\gamma}r'(r)}\frac{\partial}{\partial r}\left(K_0\left(2 e^{-\gamma}\frac{r(r)}{a}\right)\right).
\end{equation}
Equations (18) and (19), along with $\psi(r) = K_0(\frac{2e^{-\gamma}r(r)}{a})$, allow us to rewrite Eq.~(\ref{E:del1}) as 
\begin{equation} \label{E:del2}
\nabla^{2}\psi(r)=\left(\frac{r'(r)}{r}+r''(r)-\frac{r'(r)^{2}}{r(r)}\right)\frac{1}{r'(r)}\frac{\partial}{\partial r}\psi(r)+\frac{4r'(r)^{2}}{a^{2}e^{2\gamma}}\psi(r).
\end{equation}
We now make the important observation that the term in the parentheses on the right hand 
side of Eq.~(\ref{E:del2}) can be rewritten in terms of the 2D Laplacian acting on $\ln(r(r))$, 
{\it viz.,}
\begin{equation}
\frac{r'(r)}{r}+r''(r)-\frac{r'(r)^{2}}{r(r)}= r(r)\nabla^{2}(\ln(r(r)))~.
\end{equation}
Use of the expression above,  and replacing $r(r)$ by $r$, allows us to write Eq.~(\ref{E:del2}) as 
\begin{equation}
-\frac{\hbar^{2}}{M}\nabla^{2}\psi(r) +\frac{\hbar^{2}}{M}\nabla^{2}\ln(r)r\frac{\partial}{\partial r}\psi(r)= -\frac{4\hbar^2}{Ma^{2}e^{2\gamma}}\psi(r) .
\end{equation}
Finally, substituting $E=-\frac{4\hbar^2}{Ma^2e^{2\gamma}}$ and $\nabla^{2}(\ln(r))= 2\pi\delta^{2d}(\vec{r})$ (see Ref.~[\onlinecite{li}]) into Eq.~(\ref{E:del1}) yields 
\begin{equation}
-\frac{\hbar^{2}}{M}\nabla^{2}\psi(r) +\frac{2\pi\hbar^{2}}{M}\delta^{2d}(\vec{r})r\frac{\partial}{\partial r}\psi(r)= E\psi(r) .
\end{equation}
Evidently,  the pseudopotential
\begin{equation}\label{E:VPP}
V_{pp}^{2D}(r)=\frac{2\pi\hbar^{2}}{M}\delta^{2d}(\vec{r})r\frac{\partial}{\partial r},
\end{equation}
will reproduce the bound state wave function and binding energy for any zero-range
potential in two dimensions.  Notice that in contrast to 1D, the Dirac delta function is
modified by $r\partial/{\partial r}$, which yields well defined behaviour at the origin.  Indeed,
this modification is the so-called regularization operator referred to in the literature, which is
invoked to ensure self-adjoint property of the two-body
Hamiltonian.~\cite{busch,wodkiewicz,li,blume,olshanii1}  
In fact, our Eq.~(24) is simply a member of a {\em family} of 2D pseudopotentials,~\cite{olshanii2} {\em any} of
which will reproduce the shape independent results we have just obtained.
In order to clarify this point, let us consider
the 2D zero-range pseudopotential as derived by W\'odkiewicz in Ref.~[\onlinecite{wodkiewicz}], which is
a particular member of the 2D family:~\cite{author_note3}
\begin{equation}\label{E:VW}
V^W(r)=-a_2\delta^{2d}(\vec{r})\left[1-\ln\left(\sqrt{\pi}\frac{r}{L}e^{\frac{\gamma}{2}}\right)r\frac{\partial}{\partial r}\right],
\end{equation}
where $a_2$ is a coupling constant, $\gamma$ is the Euler constant and $L$ is a characteristic length.    This pseudopotential is clearly of a different form from that obtained in Eq.~(24).
We now proceed to investigate the effects of Eq.~(24) and Eq.~(\ref{E:VW}) as $r\to0^+$.  
Allowing Eq.~ (\ref{E:VPP}) to operate on 
the bound state wave function gives
\begin{equation}\label{VppSmallr}
V(r)\psi(r)=\frac{2\pi\hbar^{2}}{M}\delta^{2d}(\vec{r})r\left(-\frac{2e^{-\gamma}}{a}K_1\left(\frac{2e^{-\gamma}r}{a}\right)\right).
\end{equation}
As $r\to0^+$ the first order, modified Bessel function approaches 
$K_1\left(y\right)\rightarrow\frac{1}{y}$ and our expression in Eq.~ (\ref{VppSmallr}) reduces to
\begin{equation}
V(r)\psi(r)=-\frac{2\pi\hbar^{2}}{M}\delta^{2d}(\vec{r}).
\end{equation}
We now similarly investigate $V^W$, and easily obtain
\begin{equation} \label{E:VWWF}
V^W(r)\psi(r)=-a_2\delta^{2d}(\vec{r})\left[K_0\left(\frac{2e^{-\gamma}r}{a}\right)-\ln\left(\sqrt{\pi}\frac{r}{L}e^{\frac{\gamma}{2}}\right)r\left(-\frac{2e^{-\gamma}}{a}K_1\left(\frac{2e^{-\gamma}r}{a}\right)\right)\right].
\end{equation}
Given that $r\left(-\frac{2e^{-\gamma}}{a}K_1\left(\frac{2e^{-\gamma}r}{a}\right)\right) \to -1$ for $r\to 0^+$, we also have the relation $K_0\left(\frac{2e^{-\gamma}r}{a}\right) \to -\ln{\frac{e^{-\gamma}r}{a}}-\gamma$. These two properties allow us to write the small $r$ behaviour of (\ref{E:VWWF}) as
\begin{equation}\label{VWsmallr}
V^W(r)\psi(r)=-a_2\delta^{2d}(\vec{r})\left(\ln\left(\frac{a\sqrt{\pi}e^{\frac{\gamma}{2}}}{L}\right)\right).
\end{equation}
The coupling constant $a_2$ can now be related to our scattering length $a$ as follows.  In 
Ref.~[\onlinecite{wodkiewicz}], the bound state energy in terms of $a_2$ is given by   
\begin{equation}
E=-\frac{4\hbar^2\pi}{ML^2}e^{-\frac{4\hbar^2\pi}{Ma_2}-\gamma}.
\end{equation}
This energy must be the same as our binding energy given in terms of the scattering length, 
{\it viz.,}  $E=-\frac{4\hbar^2}{Ma^2e^{2\gamma}}$.  Equating these expressions gives us the desired 
relationship between $a_2$ and the scattering length $a$:
\begin{equation}
a_2=\frac{2\pi\hbar^2}{M}\left(\ln\left(\frac{a\sqrt{\pi}e^{\frac{\gamma}{2}}}{L}\right)\right)^{-1}.
\end{equation}
Using this expression along with (\ref{VWsmallr}) gives us
\begin{equation}
V^W(r)\psi(r)=-\frac{2\pi\hbar^{2}}{M}\delta^{2d}(\vec{r}).
\end{equation}
Consequently, we see that as $r\to0^+$ both our pseudopotential, $V(r)$, and the pseudopotential of W\'{o}dkiewicz, $V^W(r)$, are operationally equivalent over the appropriate space of wave functions.~\cite{olshanii2}  Thus,
although the pseudopotentials given by Eqs.~(24) and (25) have different forms, they
both lead to the same low-energy physics.  Indeed, in the next section, we expand upon
this result by showing that the energy spectrum of the 1D and 2D systems are
{\em independent} of the details of the interaction potential, provided the range of the potential is
much smaller than the oscillator length.

\section{Energy spectrum of the two-body problem}

\subsection{One dimensional treatment}

Consider the problem of  two identical bosons, each of mass $M$, confined by a 1D harmonic 
oscillator potential. To begin, we will first consider the two particles to be non-interacting. Each particle is subject to the potential $\frac{1}{2}Mw^2x^2$. In the center of mass, $X$, and relative coordinate, $x$, we have the Hamiltonian 
\begin{equation}
H= -\frac{\hbar^{2}}{2\mu}\frac{d^2}{dx^2} +\frac{1}{2}\mu\omega^{2}x^{2} -\frac{\hbar^{2}}{2M_c}\frac{d^2}{dX^2} +\frac{1}{2}M_c\omega^{2}X^{2},
\end{equation}
where $\mu=M/2$ is the reduced mass and $M_c=2M$ is the total mass of the system.  As we
will ultimately be interested in the two-body interaction between the atoms, we focus on the relative coordinate, where the time-independent Schr\"{o}dinger equation  is given by  
\begin{equation}\label{ODEHO1D}
-\frac{\hbar^2}{M}\frac{d^2 \psi(x)}{dx^2} +\frac{1}{4}Mw^2x^2\psi(x) = E\psi(x),
\end{equation}
and $E$ is the relative energy. Equation (\ref{ODEHO1D}) is simplified by defining $\eta=\frac{2E}{\hbar\omega}$,  $z=\frac{x}{\sqrt{2}l}$, $l^2=\frac{\hbar}{M\omega}$, which correspond
to the dimensionless energy, dimensionless length, and oscillator length, respectively.
The preceding substitutions transform Eq.~(\ref{ODEHO1D}) into 
  \begin{equation}\label{ODEHO1Dsimp}
-\frac{d^2 \psi}{dz^2} +z^2\psi = \eta \psi.
\end{equation}
We now assume the solution to (\ref{ODEHO1Dsimp}) is of the form $\psi=e^{-\frac{z^2}{2}}f(z)$, where $f(z)$ is some function of $z$. By defining $y=z^2$ and $f(z)=w(y)$ we can write $\psi= e^{-\frac{y}{2}}w(y)$, which upon substitution into Eq.~(\ref{ODEHO1Dsimp}) gives
\begin{equation}
y \frac{d^2}{dy^2}w(y) +\left(\frac{1}{2}-y\right)\frac{d}{dy}w(y)-\frac{1-\eta}{4}w(y)=0.
\end{equation}
This equation is of the confluent hypergeometric type~\cite{handbook} and as such is solved by a linear combination of confluent hypergeometric functions.  Equation (36) is of the general form 
\begin{equation}
\frac{d^2}{dy^2}v(y) +(b-y)\frac{d}{dy}v(y)-av(y)=0~,
\end{equation}
and for non-integral $b$ has the solution~\cite{handbook}
 \begin{equation}
 v(y)= c_1M(a,b,y)+c_2y^{1-b}M(a-b+1,2-b,y),
\end{equation} 
where $M$ is the confluent hypergeometric function of the first kind.~\cite{handbook} 
The solution to Eq.~(36) is therefore given by
\begin{equation}\label{CHGF1D}
w(y)=c_1M\left(\frac{1-\eta}{4},\frac{1}{2},y\right)+c_2y^{\frac{1}{2}}M\left(\frac{3-\eta}{4},\frac{3}{2},y\right).
\end{equation}
Equation (\ref{CHGF1D}), along with our prior definitions  gives us the wave function
\begin{equation}\label{HOSOL1D}
\psi(z)=\left(c_1M\left(\frac{1-\eta}{4},\frac{1}{2},z^2\right)+c_2(z^2)^{\frac{1}{2}}M\left(\frac{3-\eta}{4},\frac{3}{2},z^2\right)\right)e^{-\frac{z^2}{2}},
\end{equation}
where we recall $z=\frac{x}{\sqrt{2}l}$.  The ratio, $\frac{c_1}{c_2}$, can be extracted by investigating the large $z$ behaviour of Eq.~(\ref{HOSOL1D}).  For large argument, $y$, the behaviour of $M$ is given by 
\begin{equation}\label{largeRM}
M(p,q,y) \to \frac{\Gamma(q)}{\Gamma(p)}y^{p-q}e^y.
\end{equation}
Use of relation (\ref{largeRM}) gives us the large $z$ behaviour of our solution, Eq.~(\ref{HOSOL1D}), namely,
 \begin{equation}\label{HO1Dlarger}
\psi(z) \simeq \left(c_1\frac{\Gamma(\frac{1}{2})}{\Gamma(\frac{1-\eta}{4})}+c_2\frac{\Gamma(\frac{3}{2})}{\Gamma(\frac{3-\eta}{4})}\right)z^{-\frac{1+\eta}{2}}e^{\frac{z^2}{2}}. 
 \end{equation}
 For $\psi(z)$ not to diverge at large $z$, the term in the parentheses must vanish. Forcing this term to vanish leaves us with the relation 
 \begin{equation}
\frac{c_1}{c_2}=-\frac{\Gamma(\frac{1-\eta}{4})}{2 \Gamma(\frac{3-\eta}{4})}.
 \end{equation}

Next, we consider the same system,  except now the two particles interact via a short-range symmetric potential. We again solely concern ourselves with the relative coordinate for which
the time-independent Schr\"odinger equation reads
\begin{equation}
 -\frac{\hbar^{2}}{M}\frac{d^{2}}{dx^{2}}\psi(x) +\left(\frac{1}{4}M\omega^{2}x^{2}+V_s(x)\right)\psi(x) =E\psi(x),
 \end{equation}  
and $V_s(x)$ is a generic short-range interaction potential. 
If we now exclusively consider the region where $|x|\to0^+$ the harmonic potential vanishes.
We are then left with solving the problem of a short range interaction potential, which we will take to be of zero-range, so that the wave function is given by Equation (3).   
Note that even if the interaction
has a finite range, $b$, the harmonic potential can be ignored if 
$\mu\omega^2b^2\ll \hbar\omega$, which gives $b/l \ll 1$ for the validity of the spectrum derived
below.
In the the small $|x|$ region, we have, to within a constant
 \begin{equation}\label{smallRintcase}
 \psi(x) \sim (a-|x|).
 \end{equation}
 This solution must join smoothly with our solution to the harmonic potential problem. We therefore must investigate the small $|x|$ behaviour of equation Eq.~(\ref{HOSOL1D}) in an effort to join it smoothly with Eq.~(\ref{smallRintcase}). For small values of $y$ the confluent hypergeometric function $M(p,q,y)$ goes to unity and thus the small $z$ behaviour of  Eq.~(\ref{HOSOL1D}) is given by, 
 \begin{equation}\label{smallrHO1D}
\psi(z)\sim \left(c_1+c_2(z^2)^{\frac{1}{2}}\right).
\end{equation} 
Recalling that $(z^2)^{\frac{1}{2}}=\frac{|x|}{\sqrt{2}l}$, we can re-write Eq.~(\ref{smallrHO1D}) as
\begin{equation}\label{smallrHO1Dmod}
\psi(x)\sim \left(-\frac{c_1}{c_2}\sqrt{2}l-|x|\right).
\end{equation} 
Relating equations (\ref{smallRintcase})  and (\ref{smallrHO1Dmod}) gives us 
 \begin{equation}
\frac{a}{l}= -\frac{c_1}{c_2}\sqrt{2},
\end{equation}  
which, along with Eq.~(43) yields 
\begin{equation}\label{spectrum1D}
\frac{a}{l}= \frac{1}{\sqrt{2}}\left(\frac{\Gamma(\frac{1-\eta}{4})}{\Gamma(\frac{3-\eta}{4})}\right).
\end{equation}  
Equation (\ref{spectrum1D}) is identical to the result obtained by  Busch {\it et al.}~\cite{busch}
but has been derived here with no explicit mention of the form of the interaction. 
Therefore, the energy spectrum in 1D
is universal, and independent of the shape of the short-range interaction potential,
provided the range is much smaller than the oscillator length.

\subsection{Two dimensional treatment}

While the  procedure to determine the energy spectrum in 2D closely follows the 1D treatment,
there are additional complications associated with 2D
 which warrant further discussion.  Other than dimensionality, the system we consider is identical to that presented
 in section IIIA.  Therefore, we immediately write the 
 non-interacting Schr\"odinger
 equation in the relative coordinate as
\begin{equation} \label{E:ODEHO}
-\frac{\hbar^{2}}{M}\frac{d^{2}}{dr^{2}}u(r) +\left(\frac{1}{4}M\omega^{2}r^{2}-\frac{\hbar^2}{M}\frac{1}{4r^2}\right)u(r) =Eu(r),
\end{equation}
where $u(r)=\sqrt{r}\psi(r)$.  Upon making the same substitutions as in the 1D problem, we
obtain the following differential equation:
\begin{equation}\label{E:HOsimp}
-\frac{d^2}{dz^2}u +(z^{2} -\frac{1}{4z^2})u=\eta u.
\end{equation}
Assuming that the solution is of the form  $u=\sqrt{z}e^{-\frac{z^2}{2}}f(z)$ we can rewrite equation (\ref{E:HOsimp}) as
\begin{equation}\label{CHGF}
y \frac{d^2}{dy^2}w(y) +(1-y)\frac{d}{dy}w(y)-\frac{2-\eta}{4}w(y)=0.
\end{equation}
From a comparison with Eq.~(37), we see that this Eq.~(\ref{CHGF}) is also of the confluent 
hypergeometric kind.   The solution to Eq.~(\ref{CHGF}), is 
given by~\cite{macdonald}
\begin{equation}\label{CHGF1}
w(y)=c_1M\left(\frac{2-\eta}{4},1,y\right)+c_2W\left(\frac{2-\eta}{4},1,y\right),
\end{equation}
where $M$ and $W$ are confluent hypergeometric functions of the first and second kind,
respectively.  Equation (\ref{CHGF1}) typifies the central complication associated with the
2D case, namely, the second argument of $W$ is of {\em integral} value; in 1D, the second argument of $W$ is $1/2$.  When the second argument
of $W$ is non-integral, it may be written in terms of $M$, {\it viz.,}~\cite{macdonald}
\begin{equation}
W(a,b,y) = y^{1-b}M(a-b+1,2-b,y),
\end{equation}
which explains why Eq.~(39) can be written soley in terms of $M$.
In 2D the integral value of the second argument of $W$ does not grant us the ability use Eq.~(54).
This is not an insurmountable obstacle, as we now show.  

Our solution (\ref{CHGF1}) along with the
{\it ansatz} that $u=\sqrt{z}e^{-\frac{z^2}{2}}f(z)$, gives 
 \begin{equation}\label{HOsol2d}
u(z)= \left(c_1M\left(\frac{2-\eta}{4},1,z^2\right)+c_2W\left(\frac{2-\eta}{4},1,z^2\right)\right)\sqrt{z}e^{-\frac{z^2}{2}},
\end{equation}
where we recall $z=\frac{r}{\sqrt{2}l}$.  The ratio, $\frac{c_1}{c_2}$, is once again obtained by
investigating the large $z$ behaviour of Eq.~(\ref{HOsol2d}), where 
the functions $M$ and $W$ behave as follows,~\cite{macdonald}
\begin{equation}\label{E:largeM}
M(p,q,y) \to \frac{\Gamma(q)}{\Gamma(p)}y^{p-q}e^y~,
\end{equation}
and 
\begin{equation}\label{E:largeW}
W(p,q,y) \to \pi\cot(\pi p)\frac{\Gamma(q)}{\Gamma(p)}y^{p-q}e^y.
\end{equation}
Use of relations (\ref{E:largeM}) and (\ref{E:largeW}) gives us the large $z$ behaviour of  $u(z)$,
{\it viz.,}
\begin{equation}\label{SolHO2dlarger}
u(z) \simeq (c_1 +c_2\pi\cot(\pi p))z^{2p-3/2}e^{\frac{z^2}{2}},
\end{equation}
where $p\equiv \frac{2-\eta}{4}$.   Since Eq.~(\ref{SolHO2dlarger}) must correspond to a 
physical state, the term in the parentheses must vanish at large $z$, leaving
\begin{equation}
\frac{c_1}{c_2}= -\pi\cot(\pi p)= \tilde{\psi}(p)-\tilde{\psi}(1-p),
\end{equation}
where $\tilde{\psi}(p)$ is the digamma function and the second equality is a fundamental property of the digamma function.~\cite{handbook}

In the presence of a generic central short-range potential, $V_s(r)$,  Eq.~(\ref{E:ODEHO}) reads
 \begin{equation}\label{ODE2dinteractingHO}
 -\frac{\hbar^{2}}{M}\frac{d^{2}}{dr^{2}}u(r) +\left(\frac{1}{4}M\omega^{2}r^{2}+V_s(r)-\frac{\hbar^2}{M}\frac{1}{4r^2}\right)u(r) =Eu(r),
 \end{equation}  
where again $u(r)=\sqrt{r}\psi(r)$.  As before, we now consider the region $r\to0^{+}$, in
which  the harmonic potential vanishes leaving us with the problem of a short range potential.
For a zero range interaction (for finite range, we again require $b/l \ll 1$), 
$V_s(r)\to0$ for all $r\ne0$ and in the asymptotic region
Eq.~ (\ref{ODE2dinteractingHO}) is simply
\begin{equation}
\frac{d^{2}}{dr^{2}}u(r) +\frac{1}{4r^2}u(r) =-k^2u(r).
 \end{equation}
The solution to Eq.~(61) has already been given in Eq.~(\ref{E:WF1}), which we
recall here for convenience
\begin{equation}\label{solnonint}
u(r)= \frac{\pi}{2}\left(\cot(\delta)J_0(k r)-N_0(k r)\right)\sqrt{r}.
 \end{equation}
As we are considering the region where $r\to0^{+}$ we must explore the small $r$ behaviour of (\ref{solnonint}). For $y\to0^{+}$ the zero order Bessel and Neumann functions 
are~\cite{handbook, Adhikari}
 \begin{equation}\label{smallJ}
J_0(y)\to 1.
 \end{equation}
 and 
\begin{equation}\label{smallN}
N_0(y)\to \frac{2}{\pi}(\ln(y/2)+\gamma),
 \end{equation} 
where $\gamma$ is the Euler constant. Use of Eqs.~(\ref{smallJ}) and (\ref{smallN}) along with Eq.~(12) gives the $r\to 0^{+}$ behaviour of $u(r)$,
 \begin{equation}\label{smallrnonint}
u(r )\sim (\ln{r}-\ln{a})\sqrt{r}.
 \end{equation} 
The above solution must join smoothly with our solution for the harmonic oscillator. We therefore proceed to investigate the small $r$ behaviour of our solution of the harmonic problem, Equation (\ref{HOsol2d}). For small $r$, and hence small $z$, $M\to 1$. The problem now is with the small $r$ behaviour of 
$W$.   Specifically, what is at issue here is the lack of literature dealing with the small $r$ behaviour of $W$
when its second argument is integral.  The primary reason for this void is likely due to the fact 
that there is no
elementary relationship between $M(p,b,y)$ and $W(p,b,y)$ when $b$ is non-integral, as in {\it e.g.,} Eq.~(54).
Fortunately, in a little known paper published more than $70$ years ago, 
W.~J.~Archibald~\cite{archibald} has
developed a useful expression for $W(p,b,y)$ for integral second argument, which we
present here for $b=1$
 \begin{equation}
W(p,1,y) = M(p,1,y)(\ln(y)+\tilde{\psi}(1-p)+2\gamma)+ \sum_{n=1}^{\infty}\left(\frac{\Gamma(n+p)B_n}{\Gamma(p)\Gamma(n+1)n!}\right)y^n,
 \end{equation}  
 where $B_n= (\frac{1}{p} + \frac{1}{p+1}+...+\frac{1}{p+n-1})-2(1 + \frac{1}{2}+...+\frac{1}{n}) $, $\Gamma(\cdot)$ is the gamma function and again $\tilde{\psi}(\cdot)$ is the digamma function. From this expression it is quite evident that the small $y$ behaviour will be
\begin{equation}
W(p,1,y)\to \ln(y)+\tilde{\psi}(1-p)+2\gamma,
 \end{equation}
 and we find the small $z$ behaviour of (\ref{HOsol2d}) to be
 \begin{equation}\label{HOsmallr}
u(z)\backsimeq (c_1+c_2(\ln(z^2)+\tilde{\psi}(1-p)+2\gamma))\sqrt{z}.
\end{equation} 
Equation (\ref{HOsmallr}), upon the replacement of $z= \frac{r}{\sqrt{2}l}$ 
can be written, to within a constant,   as
 \begin{equation}\label{HOsmallr2}
u(r)\sim \left(\frac{c_1}{2c_2}+\ln(r)-\ln(l)-\frac{1}{2}\ln(2)+\frac{\tilde\psi(1-p)}{2}+\gamma\right)\sqrt{r}.
\end{equation} 
Utilizing Eqs.~ (\ref{smallrnonint}) and (\ref{HOsmallr2}) gives 
 \begin{equation}\label{spectrum1}
\frac{c_1}{2c_2}-\ln(l)-\frac{1}{2}\ln(2)+\frac{\tilde\psi(1-p)}{2}+ \gamma=-\ln(a).
\end{equation}
Equation (\ref{spectrum1}) simplifies to
 \begin{equation}
\frac{c_1}{c_2}+\tilde\psi(1-p)=\ln\left(\frac{l^2}{2 a^2}\right) + 2\ln{2} - 2\gamma,
\end{equation}
 which, upon recalling that $\frac{c_1}{c_2}=\tilde\psi(p)-\tilde\psi(1-p)$, gives the relation 
\begin{equation}
\tilde\psi(p)=\ln\left(\frac{l^2}{2 a^2}\right)+ 2\ln{2} - 2\gamma.
\end{equation}
The spectrum of the system is finally described by recalling that 
$p=-\frac{\eta}{4}+\frac{1}{2}$, whence
\begin{equation}
\tilde\psi\left(\frac{1}{2}-\frac{\eta}{4}\right)=\ln\left(\frac{l^2}{2 a^2}\right) + 2\ln{2} - 2\gamma.
\end{equation}
Equation (73) differs from the result obtained by Busch {\it et al.}~\cite{busch} by the last two terms
on the right hand side.  This difference can be traced back to the specific form for the effective range expansion,
{\it viz.,} Eq.~(12), we have used in this paper.  If we use Busch's expression\cite{author_note4}, $\cot(\delta(k)) = (2/\pi)\ln(ka)$,
the last two terms on the right hand side of Eq.~(73) disappear, in agreement with Eq.~(21) in Ref.~[\onlinecite{busch}].  Numerically, the difference between
our Eq.~(73) and Busch's Eq.~(21) is unimportant, as it would almost certainly not be resolved
in experiments.  The critical point here is that  we have shown that the 2D spectrum is independent of the
details of the short-range potential.   In addition, the 
complications arising from dimensionality are not due to the logarithmic singularities
in the pseudopotential -- as suggested in Ref.~[\onlinecite{busch}] -- but rather from the
logarithmic behaviour of $W(p,b,y)$ for integral $b=1$.   Indeed, {\em any} short-range
potential will yield the spectrum above, provided its range is much smaller than
the oscillator length.  It is nevertheless interesting to note that Eq.~(73), without specifying a form for the potential,
has naturally led to the same coupling constant, {\it viz.,} $[\ln(l^2/2a^2)]^{-1}$ which has been
used to characterize the strength of the regularized 2D zero-range interaction in earlier investigations.~\cite{busch}

\section{Conclusions}
In this paper, we examined the two-body problem of harmonically trapped ultra-cold atoms in one and
two dimensions.  We have shown that the energy spectra for both 1D and 2D is universal, in that they are independent
of the details of the short-range potential, provided the range of the potential is much less than the oscillator length.
Furthermore, we have illustrated that the concept of a zero-range pseudopotential in low-dimensional systems can
be easily understood without having to invoke the advanced mathematical language of regularization operators.
In contrast to more complicated expressions reported in the literature, we have shown that our simple 2D  zero-range
pseudopotential ({\it i.e.,} Eq.~(24), without logarithmic singularities in the potential) will yield the same
low-energy physics .~\cite{wodkiewicz,blume,olshanii1,olshanii2}  We anticipate the 1D and 2D spectra presented here to be verified by current experimental methods.~\cite{stoferle}

\begin{acknowledgments}
The authors would like to thank R. K. Bhaduri for his careful reading of the manuscript.  This work was supported
by the National Sciences and Engineering Research Council of Canada (NSERC) through the Discovery Grant program.
\end{acknowledgments}


\newpage


\begin{thebibliography}{99}
\bibitem{author_note2} To our knowledge, the universal spectrum in 1D and 2D has never
been rigorously established in the literature.  In Ref.~[\onlinecite{busch}] below, no details
are provided as to how the energy spectrum is obtained for 1D or 2D (see e.g., their Eq.~(20) and
Eq.~(21)).  The authors in Ref.~[\onlinecite{busch}] simply state that ``matters are more
complicated in two dimensions because of logarithmic singularities''.
\end{thebibliography}
\end{document}